**Gel-Confined Rolling-Circle Amplification Enables Sensitive Single-Cell Proteoform Analysis**

Luyao Zhao[a], Yanjun Yang[b], Yaochao Zheng[i], Yuhao Zhang[b], Zhengfu Huang[f], William Teng[c], Lucas Yatcyshyn[d], Jiahwei Cheong[e], Jonathan Arnold[e], Jin Xie[a], Kenan Song[f], Xianqiao Wang[f], Yiping Zhao[g], Xianyan Chen[h], Yao Yao[i], Leidong Mao[j], Yang Liu*[b, k]

[a]Department of Chemistry, The University of Georgia, Athens, Georgia 30602, USA

[b]School of Chemical, Materials and Biomedical Engineering, College of Engineering, The University of Georgia, Athens, Georgia 30602, USA

[c]Gwinnett School of Mathematics, Science, and Technology, Lawrenceville, Georgia 30044, USA

[d]Department of Bioengineering, McGill University, Montreal, Canada

[e]Complex Carbohydrate Research Center, University of Georgia, Athens, Georgia 30605, USA

[f]School of Environmental, Civil, Agricultural & Mechanical Engineering, College of Engineering, University of Georgia, Athens, Georgia 30602, USA

[g]Department of Physics and Astronomy, The University of Georgia, Athens, Georgia 30602, USA

[h]College of Public Health, The University of Georgia, Athens, Georgia 30602, USA

[i]Department of Animal & Dairy Science, College of Agricultural & Environmental Sciences, The University of Georgia, Athens, Georgia 30602, USA

[j]School of Electrical and Computer Engineering, College of Engineering, The University of Georgia, Athens, Georgia 30602, USA

[k]Institute of Bioinformatics, The University of Georgia, Athens, Georgia 30602, USA

*Email: Yang Liu (liuy@uga.edu)

## ABSTRACT

Protein abundance alone does not capture the molecular diversity generated by protein processing and post-translational modification, yet most single-cell protein assays do not resolve these proteoform states. Here, we develop RCAmp-scWB, a post-separation amplification strategy that performs rolling-circle amplification directly within the polyacrylamide gel after single-cell protein electrophoresis. Reaction–transport modeling identifies a balance between reagent access and confinement of the template-associated amplification product that supports signal amplification while retaining the electrophoretically encoded spatial readout. Using purified protein standards, RCAmp-scWB produced a 4.0-7.8 fold steeper concentration-response slope than conventional single-cell western blotting. Across six human cancer cell lines, RCAmp-scWB quantified protein abundance and resolved distinct Vimentin and PD-L1 proteoform states whose relative abundance and single-cell distributions were not captured by total protein measurements alone. Extension to individual small extracellular vesicles further revealed source-dependent shifts in Vimentin proteoform composition despite comparatively similar abundance of one major Vimentin species. By separating electrophoretic molecular discrimination from signal amplification, RCAmp-scWB enables proteoform-resolved analysis of heterogeneous cells and small extracellular vesicles.

## INTRODUCTION

Cell-to-cell variation in protein abundance and molecular state contributes to functional heterogeneity within cell populations. Protein heterogeneity extends beyond differences in abundance, as alternative splicing, proteolytic processing, and post-translational modification can generate distinct molecular forms from the same gene product, collectively described as proteoforms.[1] Protein abundance is also not reliably inferred from transcript abundance, particularly during dynamic cellular responses.[2] Single-cell protein measurements using mass cytometry, highly multiplexed imaging, oligonucleotide-barcoded antibodies, and mass spectrometry now quantify increasingly broad protein panels in individual cells.[3-10] However, most affinity-based single-cell assays identify proteins primarily through probe binding and do not independently resolve the molecular form associated with that signal. Single-cell mass spectrometry provides broader molecular coverage, but most implementations rely on bottom-up workflows in which proteins are digested before analysis, limiting direct assignment of multiple molecular features to the same intact protein molecule.[8-10] Emerging top-down approaches have begun to extend single-cell mass spectrometry toward direct intact-proteoform analysis. Single-cell Proteoform imaging Mass Spectrometry (scPiMS), for example, directly samples intact proteoforms from individual cells without prior proteolytic digestion.[11] Electrophoretic separation adds an orthogonal analytical dimension by resolving protein species before affinity-based detection. This separation dimension is particularly useful when antibodies recognize multiple molecular species or when molecular form, rather than total abundance, is the variable of interest.

Single-cell western blotting (scWB) combines microscale protein electrophoresis with subsequent immunorecognition to measure protein abundance and molecular-mass-resolved protein species in individual cells.[12-15] In scWB, proteins released from individual cells are electrophoretically separated within a thin polyacrylamide gel, immobilized at their separated positions, and subsequently detected by antibody probing.[12] Since its introduction, scWB has been extended to link whole-cell imaging with protein measurements,[13] tune electrophoretic resolution through hydrogel design,[14] resolve protein isoforms by single-cell isoelectric focusing,[16] analyze rare circulating tumor cells,[17] distinguish cytoplasmic and nuclear protein populations,[18] interrogate adherent cells without cell detachment,[19] improve probing through separation-encoded microparticles,[20] and perform three-dimensional single-cell immunoblotting.[21] More recently, we developed DropBlot, which confines single-cell lysis within droplets to retain cellular lysate and reduce analyte loss, enabling protein detection down to approximately 500 copies while extending scWB to chemically fixed cells and archived specimens.[22] A persistent limitation, however, is analytical sensitivity. Single cells contain limited quantities of many protein targets, while partitioning and diffusion of antibody probes within the polyacrylamide matrix constrain in-gel immunoprobing.[12,14,23] A recent approach addressed this limitation by transferring scWB-separated proteins from the polyacrylamide gel to nitrocellulose, where improved probe accessibility and enzyme-mediated signal amplification increased detection sensitivity.[23] Post-separation amplification improved sensitivity, but detection was moved from the original separation gel to a secondary membrane. Retaining separated proteins within the original scWB gel while amplifying signal at their immobilized positions would preserve the direct link between electrophoretic separation and molecular detection. Achieving this requires amplification without compromising the spatial information established by electrophoresis.

Rolling-circle amplification (RCA) provides isothermal nucleic-acid amplification from circular DNA templates and has been extensively adapted for localized molecular detection.[24,25] Antibody-linked DNA reporters enabled RCA to amplify protein-detection signals in immunoassays,[26] and subsequent implementations extended RCA to multiplexed protein microarrays, proximity-dependent protein assays, in situ detection of protein complexes, and microfluidic detection of cell-surface proteins.[27-30] RCA on polyacrylamide-based hydrogel antibody microarrays has demonstrated that localized amplification can be performed within a hydrogel-supported protein assay, although without a preceding electrophoretic separation dimension.[31] RCA has also been incorporated into conventional Western blotting after gel electrophoresis and membrane transfer, demonstrating that localized DNA amplification can enhance protein detection while retaining electrophoretic separation information.[32] Antibody–DNA conjugates and RCA-generated molecular barcodes have also enabled multiplexed profiling of surface proteins on individual extracellular vesicles, providing a route to quantify vesicle-to-vesicle protein heterogeneity.[33] More recently, immuno-RCA within agarose microgels enabled multiplexed single-molecule quantification of surface proteins on individual EVs.[34] These implementations establish both hydrogel-supported RCA and single-EV RCA, but do not combine localized amplification with microscale electrophoretic protein separation within the same matrix. Directly implementing RCA within the original scWB separation gel presents a distinct analytical challenge because amplification must occur at electrophoretically resolved protein bands without compromising the spatial information established during separation. In this setting, amplification becomes a coupled reaction–transport problem. DNA probes, ligase, polymerase, nucleotides, and fluorescent reporters must access antibody-bound targets within the polyacrylamide matrix, whereas the growing RCA product must remain sufficiently localized to preserve band position and width. Excessive transport restriction can suppress amplification, while insufficient confinement can broaden the amplified signal and reduce electrophoretic resolution. Effective in-gel amplification, therefore, requires sufficient reagent transport while maintaining spatial confinement of the amplified product.

Here, we develop RCAmp-scWB, a post-separation in-gel amplification strategy that performs RCA directly at electrophoretically resolved protein bands immobilized within the scWB separation gel. Following single-cell lysis, electrophoresis, and protein immobilization, target proteins are recognized with antibody–DNA conjugates, followed by padlock ligation, Phi29-mediated RCA, and fluorescent probe hybridization. We systematically optimize the in-gel RCA reaction and use reaction–diffusion modeling to examine how reagent transport and amplification-product confinement govern signal generation and spatial fidelity within the polyacrylamide matrix. We benchmark RCAmp-scWB against conventional scWB using purified protein standards and apply the approach to profile EpCAM, Vimentin, PD-L1, and HER2 across six lung and breast cancer cell lines. The electrophoretic dimension further enables single-cell analysis of Vimentin and PD-L1 proteoforms, revealing heterogeneity in both total protein abundance and proteoform composition. We further extend RCAmp-scWB to proteoform-resolved analysis of small extracellular vesicles derived from head and neck cancer cells, adding electrophoretic resolution to single-EV protein profiling. By establishing electrophoretic information before signal amplification, RCAmp-scWB separates molecular discrimination from signal generation while preserving the spatial information established during electrophoresis.

## RESULTS & DISCUSSION

### Overview and design of RCAmp-scWB

To improve protein detection in single-cell western blotting while retaining electrophoretic separation, we developed RCAmp-scWB to integrate post-separation rolling-circle amplification directly within the polyacrylamide separation gel (**Figure 1a**). The platform retains the established scWB workflow for single-cell isolation, lysis, protein electrophoresis, and photocapture (**Figure 1a,b**). Individual cells are isolated in microwells patterned within the polyacrylamide gel (PA-gel) and lysed in situ. Cellular proteins are electrophoretically separated by molecular mass and photocaptured at their migration positions using UV light, thereby preserving the spatial information generated during protein separation.

RCAmp-scWB introduces in-gel DNA-mediated amplification after protein separation and immobilization (**Figure 1c**). Photocaptured proteins are first recognized by antibody–DNA conjugates, which localize a DNA sequence to each immunorecognized protein band. A complementary padlock probe is then hybridized and ligated to form a circular DNA template, followed by primer hybridization and Phi29-mediated RCA. RCA generates a localized tandem-repeat DNA product, which is subsequently labeled through hybridization of complementary fluorescent probes. Thus, protein separation and target recognition occur before amplification, whereas RCA increases the fluorescence signal associated with the resolved protein band. In conventional scWB, by comparison, fluorescence is generated directly from the immunoprobe without a nucleic-acid amplification step. Representative protein bands remained spatially localized after RCA while exhibiting higher fluorescence intensity than those measured by conventional scWB (**Figure 1d**). Quantification revealed increased protein band intensity and signal-to-noise (S/N) ratio with RCAmp-scWB (**Figure 1e**). These results demonstrate post-separation RCA within the scWB gel while retaining the spatially resolved protein-band readout.

## Systematic optimization enhances in-gel RCA signal amplification

We next sought to establish RCA conditions that increase protein-band signal while maintaining localized detection within the PA-gel. The RCAmp-scWB amplification workflow comprises four sequential stages: padlock-probe hybridization and ligation, primer annealing, Phi29-mediated RCA, and fluorescent probe hybridization (**Figure 2a**). We systematically varied reaction conditions at each stage and quantified the resulting protein-band fluorescence to identify conditions that maximized signal amplification.

We first optimized conditions for padlock-probe hybridization and circular-template formation. Increasing padlock-probe concentration from 50 to 100 nM produced a modest increase in relative fluorescence intensity from 9.07 ± 0.43% to 10.81 ± 0.32%, whereas 200 nM increased the signal to 24.97 ± 4.31% (**Figure 2b**). Increasing the concentration further to 300 nM reduced the signal to 9.74 ± 0.91%, indicating that increased padlock-probe concentration did not monotonically increase amplification efficiency. We therefore selected 200 nM for subsequent experiments. We next varied exonuclease treatment from 30 to 120 min (**Figure 2c**). Relative fluorescence increased from 5.10 ± 0.24% at 30 min to 8.40 ± 0.81% at 60 min, but longer treatment produced no further improvement, with intensities of 7.41 ± 0.77% and 7.04 ± 0.96% at 90 and 120 min, respectively. A 60-min exonuclease treatment was therefore selected. Primer-annealing time showed a more pronounced dependence (**Figure 2d**). Increasing the incubation time from 20 to 40 min increased relative fluorescence from 4.23 ± 0.55% to 11.67 ± 1.01%, corresponding to a 2.76-fold increase. Extending annealing to 60 or 100 min reduced the signal to 6.91 ± 0.31% and 5.79 ± 0.92%, respectively, establishing 40 min as the selected primer-annealing condition.

We next optimized conditions for Phi29-mediated RCA. Among the dNTP concentrations tested, 1 mM yielded the highest relative fluorescence intensity (11.48 ± 1.21%), whereas increasing the concentration to 2, 3, or 4 mM reduced the signal to 5.30 ± 0.68%, 4.81 ± 0.23%, and 6.24 ± 0.88%, respectively (**Figure 2e**). We therefore used 1 mM dNTP for subsequent experiments. RCA reaction time also strongly affected signal generation (**Figure 2f**). Relative fluorescence increased from 1.60 ± 0.12% after 20 min to 2.93 ± 0.19% after 40 min, remained similar at 2.81 ± 0.20% after 60 min, and increased further to 6.37 ± 0.39% after 120 min. Thus, extending the reaction from 20 to 120 min produced an approximately 3.98-fold increase in relative fluorescence. Representative fluorescence images showed the corresponding accumulation of RCA-associated signal while the amplified protein remained detectable as a spatially localized band (**Figure 2g**). The longest tested reaction time, 120 min, was therefore used in the optimized workflow.

We subsequently optimized fluorescent-probe hybridization to the RCA product. Increasing fluorescent-probe concentration from 200 to 400 nM increased relative fluorescence from 20.38 ± 4.41% to 37.31 ± 4.20%, with a similar signal at 600 nM (37.60 ± 3.56%) (**Figure 2h**). Increasing the concentration to 800 nM further increased the signal to 52.70 ± 1.60%, corresponding to a 2.59-fold increase relative to 200 nM. We therefore selected 800 nM fluorescent probe. In contrast, extending probe-hybridization time did not improve detection (**Figure 2i**). The highest relative fluorescence was observed after 30 min (8.46 ± 1.31%) and progressively decreased to 6.48 ± 0.29%, 3.59 ± 0.10%, and 2.90 ± 0.49% after 60, 90, and 120 min, respectively. A 30-min probe-hybridization step was therefore used for subsequent RCAmp-scWB measurements. Collectively, these parameter sweeps showed that increasing reagent concentration or incubation time did not uniformly increase RCAmp-scWB signal, and instead identified defined operating conditions for each stage of the amplification reaction.

We then compared conventional scWB with RCAmp-scWB before and after reaction optimization (**Figure 2j,k**). Representative fluorescence images showed that optimization substantially increased the RCAmp-scWB signal relative to both conventional scWB and the baseline RCAmp-scWB condition (**Figure 2j**). This trend was further supported by summary analysis of protein-band fluorescence intensity and signal-to-noise ratio (**Figure 2k**). Importantly, the amplified readout remained spatially localized after RCA. Based on these optimization studies, the final RCAmp-scWB conditions were set to 200 nM padlock probe, 60 min exonuclease treatment, 40 min primer hybridization, 1 mM dNTP, 120 min RCA, and 800 nM fluorescent probe with a 30 min probe-incubation period. The optimized conditions were therefore used for subsequent analytical characterization and single-cell measurements. Although these experiments established conditions that increased fluorescence output, signal enhancement alone does not determine whether RCA preserves the spatial information encoded by electrophoretic separation. We next examined how transport of RCA components and confinement of the amplification product within the PA-gel influence signal localization.

**Reaction–transport modeling reveals selective reagent access and RCA-product confinement**

Figure 2 established reaction conditions that increased RCAmp-scWB fluorescence, but amplification within the original separation gel is useful only if soluble reagents can access immobilized targets without allowing the amplified product to erode the spatial information established by electrophoresis. To examine this balance, we developed a two-dimensional reaction-transport model that describes reagent entry, RCA product generation, and fluorescent probe readout within an 8%T, 50-μm-thick PA gel (**Figure 3a**). The model resolves the electrophoretic separation direction and gel depth, with the protein-associated circular DNA template immobilized at the position established by electrophoresis. Phi29 polymerase, dNTPs, and fluorescent probes enter from the reagent-facing surface, whereas the RCA concatemer remains associated with the immobilized template. Because molecular access to PA-gels depends independently on intragel diffusivity and equilibrium partitioning, these two contributions were modeled separately, consistent with previous measurements showing strong size-dependent diffusion and partitioning of macromolecules in PA matrices.[35-37]

The model predicted a pronounced molecular-size dependence of reagent transport in the 8%T PA-gel (**Figure 3b**). For dNTPs, fluorescent probes, and Phi29 polymerase, represented by hydrodynamic radii of 0.5, 1.8, and 4.0 nm, respectively, the predicted $D_{gel}/D_0$ decreased from 0.482 to 0.212 and 0.083. The corresponding partition coefficients decreased more strongly, from 0.864 for dNTPs to 0.495 for fluorescent probes and 0.075 for Phi29. Combining diffusion and partitioning as the normalized permeability-like quantity $K$ ($D_{gel}/D_0$) yielded values of 0.417, 0.105, and 0.0062, respectively. Thus, the predicted permeability-like value for Phi29 was approximately 67-fold lower than that for dNTPs. These calculations identify macromolecular partitioning, rather than a uniform reduction in molecular diffusion, as a major physical constraint on reagent access within the PA-gel.

Strong Phi29 exclusion did not, however, imply prohibitively slow transport across the thin gel. At the deepest position, 50 μm from the solution interface, normalized Phi29 concentration reached approximately 0.16 of its intragel equilibrium value after 1 min, approximately 0.8 after 5 min, and approached equilibrium after 20 min (**Figure 3c**). The distinction between equilibration and loading is important: these profiles are normalized to the gel-side equilibrium concentration, whereas the Phi29 partition coefficient of 0.075 predicts an equilibrium intragel concentration of only approximately 7.5% of the external concentration. The model therefore predicts that Phi29 can traverse the 50-μm gel on a timescale shorter than the 120-min RCA reaction while remaining strongly depleted relative to the bulk solution. This separates transient through-thickness diffusion from the persistent partitioning penalty and suggests that diffusion distance alone is insufficient to explain the experimentally observed RCA-time dependence.

We next asked whether RCA product accumulation would compromise the lateral information encoded by electrophoresis. Because amplification initiates from an immobilized protein-associated template, the baseline model treats translational mobility of the growing concatemer as negligible. Under this tethered-product condition, the predicted RCA signal remained spatially registered with the original electrophoretic band after 120 min (**Figure 3d**). The initial band FWHM was 38.35 μm and remained 38.35 μm after amplification, corresponding to an FWHM ratio of 1.00 (**Figure 3e**). The nearly overlapping pre-amplification and predicted RCA fluorescence profiles therefore show that substantial signal generation does not intrinsically require lateral redistribution of the amplified product. In this model, spatial preservation arises from the combination of an immobilized template and localized concatemer growth rather than from restricting transport of the soluble RCA reagents.

We then varied PA-gel density to determine how network structure shifts the balance between reagent access and product confinement (**Figure 3f**). Increasing gel density from 5%T to 12%T decreased the effective mesh-size parameter from 287 to 20.7 nm. Over the same range, the predicted Phi29 $D_{gel}/D_0$ decreased from 0.202 to 0.026, while its partition coefficient decreased from 0.404 to 0.0079. Consequently, the Phi29 permeability-like metric decreased from 0.0816 to $2.06\times10^{-4}$, an approximately 400-fold reduction. Predicted endpoint fluorescence decreased correspondingly to 2.3% of the 5%T value at 12%T. In contrast, the mean confinement index $R_g/\zeta$ increased from 0.84 at 5%T to 1.67 at the experimental 8%T condition and approached approximately 1.9 at higher gel densities. At 8%T, the model therefore enters a regime in which the effective RCA product size exceeds the gel structural length scale ($R_g/\zeta>1$), while the predicted endpoint signal remains approximately 22% of that obtained at 5%T. These simulations expose an opposing dependence on gel density: increasing polymer content strengthens product confinement but rapidly reduces macromolecular access and amplification output. The full gel-structure sweep therefore identifies a transport-confinement trade-off rather than a simple benefit of increasing or decreasing gel density.

The temporal simulation further linked RCA accumulation to the progressive development of polymer confinement (**Figure 3g**). Predicted signal reached 0.15, 0.33, and 0.51 of the 120-min endpoint after 20, 40, and 60 min, respectively. Over the same interval, the mean confinement index increased from 0.63 at 20 min to 0.94 at 40 min and 1.17 at 60 min, before reaching 1.67 at 120 min. The model, therefore, places the transition to network-scale confinement ($R_g/\zeta$) between approximately 40 and 60 min, while RCA product continues to accumulate. Experimental RCA-time measurements showed the same overall increase toward the 120-min endpoint, although the experimental signal exhibited a plateau between 40 and 60 min that was not reproduced by the monotonic simulation. The model should therefore be interpreted as capturing the first-order transport and confinement behavior rather than all assay-specific intermediate kinetics.

Overall, the model identifies a physical separation between reagent transport and amplified-product localization in RCAmp-scWB. Small substrates remain comparatively accessible to the PA-gel, Phi29 experiences substantial partitioning but can penetrate the thin gel within the experimental reaction time, and the template-associated RCA product remains spatially confined as amplification proceeds. This combination provides a mechanistic basis for increasing fluorescence signal while retaining the electrophoretic position and width of the original protein band.

**Analytical characterization of RCAmp-scWB using purified protein standards**

Having established the RCA reaction conditions and the reaction-transport framework governing in-gel amplification, we next quantified the analytical response of RCAmp-scWB using purified BSA488 as a protein standard. BSA488 was analyzed at concentrations from 1.5 to 1500 nM, spanning a 1,000-fold concentration range, and directly compared with conventional scWB **(Figure 4a,b)**. RCAmp-scWB produced higher fluorescence intensities than conventional scWB at each concentration tested. Mean fluorescence increased from 2.88 ± 0.44 a.u. at 1.5 nM to 13.69 ± 1.53, 33.92 ± 3.30, and 60.74 ± 5.75 a.u. at 15, 150, and 1500 nM, respectively **(Figure 4c)**. Under the same conditions, conventional scWB yielded 0.69 ± 0.16, 5.35 ± 0.90, 8.74 ± 1.10, and 15.82 ± 1.52 a.u., respectively. Thus, post-separation RCA increased the measured fluorescence signal throughout the tested concentration range while retaining a concentration-dependent response.

To quantify the concentration response independently of absolute signal magnitude, fluorescence intensity was regressed against log10-transformed BSA488 concentration for each independent device. The calibration fits were consistent across devices, with $R^2$ values of 0.963–0.972 for RCAmp-scWB and 0.967–0.984 for conventional scWB. RCAmp-scWB yielded a mean calibration slope of 19.38 ± 1.77 a.u. decade$^{-1}$, compared with 4.88 ± 0.44 a.u. decade$^{-1}$ for conventional scWB **(Figure 4d)**. All three paired measurements showed a similar increase, with RCAmp-scWB/scWB slope ratios ranging from 3.94 to 4.03, resulting in a 3.97-fold higher mean calibration slope (paired two-sided t-test, $p = 0.0028$). RCAmp-scWB therefore produced a steeper concentration-dependent fluorescence response than conventional scWB. We next examined the amplification gain at each BSA488 concentration **(Figure 4e)**. The mean RCAmp-scWB/scWB fluorescence-intensity ratio was 4.26 ± 0.62 at 1.5 nM, 2.57 ± 0.15 at 15 nM, 3.89 ± 0.14 at 150 nM, and 3.84 ± 0.07 at 1500 nM. Although the magnitude of signal amplification varied with concentration, RCAmp-scWB increased mean fluorescence by more than 2.5-fold at every concentration tested. These measurements establish reproducible signal amplification across a broad protein concentration range while preserving a quantitative concentration-dependent readout, providing the analytical basis for subsequent single-cell protein and proteoform measurements.

**Application of RCAmp-scWB to single-cell protein profiling across human cancer cell lines**

Following analytical characterization with purified protein standards, we applied RCAmp-scWB to single-cell protein profiling across six human cancer cell lines. Three lung cancer cell lines (H1299, H3122, and A549) and three breast cancer cell lines (HCC1806, MCF7, and MDA-MB-231) were analyzed for Vimentin, PD-L1, HER2, and EpCAM (**Figure 5a**). RCAmp-scWB detected all four targets at the single-cell level, enabling comparison of protein abundance across cell lines and analysis of pairwise protein co-expression states. The four proteins exhibited distinct abundance patterns across the six cell lines (**Figure 5b–e**). Vimentin fluorescence was highest in MDA-MB-231 (28,798 ± 1,251 a.u.) and H1299 (24,383 ± 783 a.u.) and lowest in MCF7 (3,925 ± 856 a.u.). PD-L1 was highest in H1299 (34,449 ± 928 a.u.), followed by HCC1806 (24,168 ± 601 a.u.) and MDA-MB-231 (19,306 ± 876 a.u.), whereas MCF7 showed the lowest mean PD-L1 signal (4,025 ± 689 a.u.). HER2 displayed a different pattern, with higher fluorescence in MCF7 (11,629 ± 459 a.u.) and H3122 (7,273 ± 988 a.u.) and lower signals in HCC1806 (1,685 ± 565 a.u.) and MDA-MB-231 (1,690 ± 555 a.u.). EpCAM was similarly highest in MCF7 (22,490 ± 151 a.u.) and H3122 (18,012 ± 450 a.u.) and lowest in MDA-MB-231 (3,206 ± 460 a.u.). Depending on the protein, the highest and lowest cell-line means differed by approximately 6.9- to 8.6-fold, demonstrating substantial cell-line-dependent variation in protein abundance.

Single-cell measurements further revealed distinct co-expression states that were not apparent from population-level protein intensities alone. For EpCAM and Vimentin, H1299 cells were predominantly classified as Vimentin-positive/EpCAM-negative (85.4%), whereas all pooled MDA-MB-231 cells fell within this quadrant (100%) (**Figure 5f**). In contrast, MCF7 cells were predominantly EpCAM-positive/Vimentin-negative (98.9%). H3122, A549, and HCC1806 contained larger double-positive populations of 28.1%, 22.5%, and 33.1%, respectively. Device-level analysis yielded mean double-positive fractions of 28.0 ± 1.4% for H3122, 21.8 ± 6.0% for A549, and 33.3 ± 1.8% for HCC1806, compared with 7.1 ± 2.8% for H1299, 1.1 ± 0.2% for MCF7, and 0% for MDA-MB-231 (**Figure 5h,i**).

HER2 and PD-L1 produced a different set of single-cell states (**Figure 5g**). H1299 contained a large HER2-positive/PD-L1-positive population (59.9%) together with a PD-L1-positive/HER2-negative population (40.1%). H3122 and MCF7 were dominated by HER2-positive/PD-L1-negative cells (99.1% and 98.4%, respectively), whereas HCC1806 was predominantly PD-L1-positive/HER2-negative (93.7%). A549 was primarily HER2-positive/PD-L1-negative (87.8%) but contained a 12.2% double-positive population. MDA-MB-231 was predominantly PD-L1-positive/HER2-negative (80.5%), with smaller double-positive (9.0%) and double-negative (9.2%) populations. Device-level double-positive fractions were highest in H1299 (59.3 ± 8.6%), followed by A549 (11.9 ± 4.3%), MDA-MB-231 (8.8 ± 4.7%), and HCC1806 (5.8 ± 3.6%); double-positive cells were not detected in H3122 or MCF7.

Vimentin/PD-L1 co-expression provided a third pattern of single-cell heterogeneity (**Figure 5h,i**). H1299 was predominantly double-positive (96.9% of pooled cells), whereas MCF7 was almost entirely double-negative (99.1%). H3122 and A549 were dominated by double-negative cells (68.7% and 77.4%, respectively), while HCC1806 consisted primarily of PD-L1-positive/Vimentin-negative (65.9%) and double-positive (33.9%) cells. MDA-MB-231 contained a large double-positive population in the pooled single-cell analysis (83.0%), although the device-level double-positive fraction showed substantial variability (61.4 ± 48.9%, n = 2 devices). HCC1806 yielded a device-level double-positive fraction of 36.5 ± 6.0% (n = 2 devices), whereas the remaining cell lines were measured across three independent devices. These pairwise measurements demonstrate that RCAmp-scWB can resolve heterogeneous protein co-expression states within and across cancer cell populations. The cell-line measurements extend RCAmp-scWB from purified-protein characterization to heterogeneous single-cell samples and establish a protein-level reference for the subsequent analysis of electrophoretically resolved Vimentin and PD-L1 proteoforms.

**Proteoform-resolved single-cell analysis of Vimentin and PD-L1**

We next used the electrophoretic separation dimension of RCAmp-scWB to resolve distinct apparent-molecular-mass species of Vimentin and PD-L1 at the single-cell level. For each protein, two reproducibly separated species were detected and designated VIM′ and VIM″ for Vimentin and PD-L1′ and PD-L1″ for PD-L1, with the prime and double-prime species corresponding to the higher- and lower-apparent-molecular-mass forms, respectively (**Figure 6a,b**). Across pooled protein bands, VIM′ and VIM″ migrated to mean positions of 118.1 ± 27.1 μm and 269.8 ± 44.9 μm, respectively, corresponding to an approximately 152-μm separation, whereas PD-L1′ and PD-L1″ migrated to 251.1 ± 24.1 μm and 459.7 ± 31.7 μm, corresponding to an approximately 209-μm separation. The observation of mobility-resolved species is consistent with the extensive post-translational processing reported for both proteins. PD-L1 is strongly N-glycosylated, with glycosylation shifting the apparent molecular mass of the approximately 33-kDa polypeptide to heterogeneous species commonly observed at approximately 45–55 kDa.[38] PD-L1 can also undergo mono- and multiubiquitination, generating additional higher-molecular-mass species.[39] Vimentin is likewise regulated by multiple post-translational modifications, including phosphorylation, glycosylation, SUMOylation, ubiquitination, acetylation, and citrullination, several of which alter filament organization, solubility, or electrophoretic behavior.[40,41] We therefore interpret VIM′/VIM″ and PD-L1′/PD-L1″ as electrophoretically resolved proteoform states without assigning either species to a specific biochemical modification.

Proteoform abundance varied substantially across the six cancer cell lines (**Figure 6c,d**). VIM′ was most abundant in MDA-MB-231 (6,219 ± 604 a.u.) and H1299 (5,672 ± 278 a.u.) and lowest in MCF7 (608 ± 277 a.u.), whereas VIM″ was highest in MDA-MB-231 (22,579 ± 650 a.u.) and H1299 (18,711 ± 522 a.u.) and lowest in MCF7 (3,316 ± 589 a.u.). VIM″ accounted for approximately 76.7–89.5% of the total Vimentin signal across the six cell lines, indicating that variation in total Vimentin abundance reflected both overall protein level and the relative contribution of the resolved species. PD-L1 showed a different proteoform composition. PD-L1′ was the predominant species in all six cell lines, contributing approximately 83.3–91.8% of total PD-L1 signal. PD-L1′ was highest in H1299 (29,319 ± 601 a.u.) and HCC1806 (21,104 ± 553 a.u.) and lowest in MCF7 (3,493 ± 541 a.u.), whereas PD-L1″ remained substantially lower than PD-L1′ across all six cell lines. Thus, the relative predominance of PD-L1′ was conserved despite substantial cell-line-dependent variation in total PD-L1 abundance.

Single-cell analysis revealed additional heterogeneity that was not apparent from device-level mean abundance alone (**Figure 6e,f**). Among pooled H1299 and MDA-MB-231 cells, 79.7% and 89.2%, respectively, were positive for both Vimentin proteoforms. In contrast, A549 was dominated by a VIM′−/VIM″+ state (62.6%), with only 30.5% of cells positive for both species, whereas H3122 and MCF7 contained large double-negative populations (57.9% and 78.8%, respectively). HCC1806 showed a broader distribution across the four Vimentin proteoform states, comprising 31.6% VIM′+/VIM″+, 31.1% VIM′+/VIM″−, 9.9% VIM′−/VIM″+, and 27.5% VIM′−/VIM″− cells. PD-L1 proteoforms produced a different pattern: H1299 cells were predominantly PD-L1′+/PD-L1″+ (81.1%), with large double-positive populations also observed in HCC1806 (65.7%) and MDA-MB-231 (66.5%). MDA-MB-231 additionally contained a substantial PD-L1′+/PD-L1″− population (31.4%), whereas A549, H3122, and MCF7 were dominated by cells negative for both resolved PD-L1 species (74.7%, 80.4%, and 84.7%, respectively). These distributions demonstrate that similar population-level protein measurements can arise from substantially different combinations of proteoform states at the single-cell level.

An AUC-based heatmap integrating total protein abundance with the resolved Vimentin and PD-L1 species further revealed cell-line-specific molecular signatures (**Figure 6g**). H1299 showed elevated Vimentin and PD-L1 signals across both total-protein and proteoform measurements, whereas MCF7 was characterized by higher HER2 and EpCAM and lower Vimentin and PD-L1 features. MDA-MB-231 displayed a strong Vimentin signature, particularly for VIM″. Importantly, the proteoform measurements did not simply reproduce differences in total protein abundance: cell lines with similar total-protein levels could differ in the relative abundance and single-cell distribution of the resolved species. To determine whether proteoform composition contributed information beyond total protein abundance, we performed principal component analysis using matched single-cell measurements. PCA based on total Vimentin, PD-L1, HER2, and EpCAM abundance showed that PC1 and PC2 accounted for 71.7% and 11.7% of the total variance, respectively (**Figure 6h**). PC1 was driven by positive contributions from Vimentin and PD-L1 and opposing contributions from HER2 and EpCAM, consistent with the major protein-expression differences observed across the six cell lines. When VIM′ fraction and PD-L1′ fraction were added to the same single-cell feature set, PC1 and PC2 accounted for 50.0% and 17.4% of the variance, respectively (**Figure 6i**). The expanded feature space altered the distribution of single cells and redistributed variance across the principal components. Analysis of the corresponding loadings showed that total Vimentin, PD-L1, HER2, and EpCAM remained the major contributors to PC1, with loadings of 0.503, 0.485, −0.454, and −0.492, respectively, whereas PD-L1′ fraction showed the largest absolute loading on PC2 (−0.920) (**Figure 6j**). VIM′ fraction contributed more strongly to PC3 (0.855) than to PC1 or PC2, indicating that proteoform composition introduced variation that was not captured by total protein abundance alone.

Protein abundance and proteoform composition therefore represent separable but complementary dimensions of single-cell heterogeneity. The matched single-cell PCA further showed that proteoform fractions contributed variance along dimensions distinct from those dominated by total protein abundance. RCAmp-scWB thus extends single-cell protein analysis beyond quantifying how much of a target protein is present to resolving the molecular forms that contribute to that signal. The biochemical identities of VIM′/VIM″ and PD-L1′/PD-L1″ remain to be established, and targeted deglycosylation, phosphatase treatment, modification-specific immunoprobing, or proteoform-resolved mass spectrometry could provide direct biochemical assignment of the resolved species.

**Proteoform-resolved profiling of individual small extracellular vesicles**

Following single-cell proteoform analysis, we next asked whether RCAmp-scWB could resolve proteoform heterogeneity at the level of individual small extracellular vesicles (sEVs). sEVs derived from the head and neck cancer cell lines HN6 and HN12 were analyzed for the two electrophoretically resolved Vimentin species, VIM′ and VIM″ (**Figure 7a**). Across individual-sEV measurements, the two species occupied distinct migration regions in both sEV populations (**Figure 7b**). VIM′ migrated to 44.4 ± 0.4 µm in HN6-derived sEVs and 47.0 ± 0.3 µm in HN12-derived sEVs, whereas VIM″ migrated to 88.0 ± 0.3 µm and 87.3 ± 0.3 µm, respectively. The preserved separation between VIM′ and VIM″ across the two sEV sources supports detection of the same two electrophoretically resolved Vimentin proteoform states in both populations.The two proteoforms showed markedly different abundance patterns between sEV sources (**Figure 7c**). VIM″ fluorescence was nearly unchanged between HN6- and HN12-derived sEVs, with mean intensities of 3868.8 ± 293.0 a.u. and 3905.6 ± 188.7 a.u., respectively. In contrast, VIM′ increased from 366.8 ± 59.8 a.u. in HN6-derived sEVs to 882.7 ± 70.8 a.u. in HN12-derived sEVs, corresponding to a 2.41-fold increase. Consequently, the combined VIM′ and VIM″ fluorescence differed by only approximately 13% between the two sEV populations, whereas the VIM′ component alone differed by more than twofold. The molecular distinction between HN6- and HN12-derived sEVs was therefore driven primarily by the higher-apparent-molecular-mass VIM′ species rather than by a uniform increase in both Vimentin proteoforms.

Single-sEV measurements further showed that this difference reflected a pronounced redistribution of proteoform states (**Figure 7d–f**). Among pooled HN6-derived sEVs, 30.5% were negative for both proteoforms, 30.6% were VIM″+/VIM′−, 15.4% were VIM″−/VIM′+, and 23.5% were positive for both VIM′ and VIM″. HN12-derived sEVs showed a substantially different composition: only 5.7% were double-negative and 4.0% were VIM″+/VIM′−, whereas 35.9% were VIM″−/VIM′+ and 54.4% were double-positive. At the device level, the VIM′+/VIM″+ fraction increased from 23.5 ± 4.5% in HN6-derived sEVs to 54.4 ± 2.0% in HN12-derived sEVs (**Figure 7e**). The fraction of VIM′-positive sEVs likewise increased from 38.8 ± 7.2% to 90.3 ± 0.2%, whereas the VIM″-positive fraction changed only modestly, from 54.1 ± 1.1% to 58.4 ± 2.5%.

Thus, the difference between the two sEV populations was dominated by redistribution into VIM′-containing proteoform states rather than by a broad increase in Vimentin-positive sEVs. These measurements extend RCAmp-scWB from intact cells to individual sEVs and demonstrate that electrophoretic separation can preserve molecular-state information in samples with limited analyte content. Importantly, treating Vimentin as a single affinity-defined target would substantially obscure the distinction between the HN6- and HN12-derived sEV populations: VIM″ abundance and prevalence were similar, whereas VIM′ abundance and the frequency of VIM′-containing sEV states differed markedly. RCAmp-scWB therefore extends proteoform-resolved single-cell analysis to the single-sEV level, revealing molecular heterogeneity in the forms contributing to an apparently similar total protein signal.

RCAmp-scWB establishes a post-separation amplification strategy in which electrophoretic protein information is defined before signal amplification and retained within the original PA-gel. Reaction–transport analysis showed that this architecture is enabled by a balance between reagent access and confinement of the template-associated RCA product, supporting signal amplification while preserving the electrophoretically encoded spatial readout. Across human cancer cell lines, RCAmp-scWB resolved not only differences in protein abundance and co-expression but also distinct Vimentin and PD-L1 proteoform states whose relative abundance and single-cell distributions were not captured by total protein measurements alone. Extension to individual sEVs further demonstrated that proteoform composition can reveal molecular-state differences that remain obscured when a protein is treated as a single affinity-defined target. By separating molecular discrimination from signal amplification, RCAmp-scWB expands single-cell western blotting from protein-abundance measurement to proteoform-resolved analysis of heterogeneous cells and sEVs.

## METHODS

**Computational modeling of gel-confined RCA.** A two-dimensional reaction–transport model was developed to describe RCA within the PA-gel along the electrophoretic separation axis and gel depth. Phi29 polymerase, dNTPs, and fluorescent probes were modeled as diffusing from the reagent-facing surface, with molecular access determined by both intragel diffusion and equilibrium partitioning. The protein-associated RCA template was treated as immobilized, and the growing RCA concatemer was assumed to remain spatially localized at the template position. Phi29-mediated amplification was described using literature-informed reaction kinetics, followed by fluorescent-probe transport and binding to the RCA product. Product confinement was characterized by the ratio of the effective RCA-product radius of gyration to the PA-gel mesh size. For gel-density sensitivity analysis, transport, partitioning, effective mesh size, RCA product accumulation, and fluorescence output were recalculated for each simulated %T condition. Simulations were implemented in MATLAB; detailed equations, parameter definitions, and sensitivity analyses are provided in the Supplementary Methods.

**Fabrication of microwell-arrayed polyacrylamide gel slides**. Microwell-arrayed polyacrylamide (PA) gel slides for single-cell Western blotting (scWB) were fabricated using an SU-8-patterned silicon mold. An 8%T polyacrylamide precursor solution was prepared using 30% acrylamide/bis-acrylamide solution (29:1; Sigma-Aldrich, A3574), benzophenone methacrylamide (BP-APMA; 3-benzoyl-N-[3-(2-methyl-acryloylamino)-propyl]benzamide, CAS 1706951-11-2; Raybow USA, Brevard, NC, USA), 1× Tris–glycine buffer (Sigma-Aldrich, T4904), and deionized water. BP-APMA was incorporated into the gel matrix as a photoreactive capture monomer to enable subsequent UV-mediated immobilization of electrophoretically separated proteins.

The gel precursor solution was degassed prior to polymerization. Ammonium persulfate (APS; Sigma-Aldrich, A3678) and N,N,N′,N′-tetramethylethylenediamine (TEMED; Sigma-Aldrich, T9281) were then added to final concentrations of 0.08% (w/v) and 0.08% (v/v). The precursor solution was immediately cast between the patterned silicon mold and a silanized glass slide and allowed to polymerize for approximately 20 min at room temperature.Following polymerization, the PA-gel-coated glass slide was carefully released from the mold to preserve the microwell array. The fabricated gels were immersed in deionized water to remove residual unpolymerized components and were stored hydrated in deionized water until use.

**Cell culture.** Human non-small cell lung cancer (NSCLC) cell lines H1299, H3122, and A549 and human breast cancer cell lines MDA-MB-231, MCF7, and HCC1806 were obtained from the American Type Culture Collection (ATCC, Manassas, VA, USA). Human head and neck cancer cell lines HN6 and HN12 were kindly provided by Dr. Yong Teng (Emory University, Atlanta, GA, USA). All cell lines were used for experiments before passage 10 and maintained in Dulbecco's Modified Eagle Medium (DMEM, high glucose; Thermo Fisher Scientific, 11965092) or RPMI-1640 medium (Thermo Fisher Scientific, 11875093), as appropriate for each cell line. Culture media were supplemented with 10% (v/v) fetal bovine serum (FBS; Thermo Fisher Scientific, A5670701), 0.1 mM non-essential amino acids (NEAA; Thermo Fisher Scientific, 11140050), and 1% (v/v) penicillin–streptomycin (Thermo Fisher Scientific, 15140122). Cells were maintained at 37 °C in a humidified incubator with 5% $CO_2$. For RCAmp-scWB experiments, adherent cells were dissociated using 0.05% or 0.25% trypsin–EDTA (Thermo Fisher Scientific, 25300054 or 25200072) at 37 °C for approximately 5 min. Cells were collected by centrifugation at 300 × g for 3 min and resuspended in phosphate-buffered saline (PBS) to approximately $1 \times 10^6$ cells mL$^{-1}$. Cell concentrations were determined using an automated cell counter (Countess, Thermo Fisher Scientific).

**Single-cell loading, cell lysis, electrophoresis, and protein photocapture.** Microwell-arrayed PA-gel slides were equilibrated in PBS before cell loading. A cell suspension at approximately $1 \times 10^6$ cells mL$^{-1}$ was applied to the surface of the microwell array and allowed to settle for approximately 10 min to facilitate gravitational loading of individual cells into the microwells. Excess cells remaining on the gel surface were removed by gentle rinsing with PBS to obtain single-cell occupancy. Following cell loading, the gel was transferred to the electrophoresis apparatus and contacted with preheated lysis buffer (RIPA like) at 55 °C. The lysis buffer contained sodium dodecyl sulfate (SDS), sodium deoxycholate, Triton X-100, and Tris–glycine buffer. Cells were lysed in situ for 30 s, after which proteins were electrophoretically separated under an electric field of 240 V cm$^{-1}$ for 15 s. Immediately following electrophoresis, separated proteins were immobilized within the BP-APMA-containing gel by UV-mediated photocapture. Gels were irradiated at 350–360 nm with an approximate exposure of 1.8 J cm$^{-2}$ for 45 s. Following photocapture, gels were extensively washed with Tris-buffered saline containing Tween-20 (TBST) 30min to remove residual electrophoresis buffer, detergents, and unbound cellular components before immunodetection.

**Conventional single-cell Western blot immunoprobing.** For conventional fluorescence-based scWB detection, primary antibodies were diluted 1:50 in 2% (w/v) bovine serum albumin (BSA) in TBST and incubated with the gel for 2 h at room temperature or overnight at 4C. Following primary antibody incubation, gels were washed 60min for three times with TBST to remove unbound antibody. Fluorophore-conjugated secondary antibodies were diluted 1:100 in 2% (w/v) bovine serum albumin (BSA) in TBST and incubated with the gels for 1 h at room temperature.

Gels were subsequently washed 60min for three times with TBST, desalted in deionized water, and dried under a gentle stream of nitrogen before fluorescence imaging. For comparison with the RCA-amplified detection workflow, conventional scWB samples were prepared from the same protein-separation and photocapture procedure and were imaged using identical acquisition settings whenever direct fluorescence-intensity comparisons were performed.

**Design of antibody-linked DNA barcodes and RCA oligonucleotides.** Each target-specific antibody was conjugated to a 5′-thiol-modified single-stranded DNA (ssDNA) barcode. The barcode sequence contained a target-recognition region complementary to the terminal arms of the corresponding padlock probe and provided the molecular template required for subsequent padlock circularization.Padlock probes were synthesized with a 5′ phosphate to permit enzymatic ligation after hybridization to the antibody-linked DNA barcode. Following circularization, circularized padlock sequence was used as template for amplification by phi29 DNA polymerase. Fluorescently labeled oligonucleotide probes complementary to the repetitive RCA product were subsequently hybridized to the amplified DNA concatemer to generate the fluorescence readout. DNA barcodes were designed to minimize sequence complementarity between different assay components and to reduce nonspecific hybridization. Oligonucleotide sequences used for antibody conjugation, padlock ligation, RCA priming, and fluorescent detection are ordered in IDT provided in Supplementary Table S1.

**Antibody–DNA conjugation**. For each conjugation reaction, 20 µg of antibody was reacted with 1 µL of 4 mM sulfosuccinimidyl 4-(N-maleimidomethyl)cyclohexane-1-carboxylate (Sulfo-SMCC) prepared in dimethyl sulfoxide (DMSO). The reaction was incubated for 2 h at room temperature. After 1h, 5′-thiol-modified DNA oligonucleotides at 100 µM were incubated with 100 mM dithiothreitol (DTT) in PBS containing 5 mM EDTA for 1 h at 37 °C. Activated antibodies and reduced oligonucleotides were purified separately using Zeba Spin Desalting Plates, 7 K MWCO (Thermo Scientific) to remove excess Sulfo-SMCC and DTT. Purified maleimide-activated antibodies were combined with the reduced thiol-modified DNA oligonucleotides and allowed to conjugate 1h at room temperature and directly followed by dialysis in a Slide-A-Lyzer MINI Dialysis Device, 7 K MWCO, 0.1 ml (Thermo Scientific) against 100mL PBS with constant stirring by a magnetic bar at 4 °C overnight. The resulting antibody–DNA conjugates were adjusted to an antibody concentration of approximately 1 μM in PBS containing 0.1% (w/v) BSA and stored at 4 °C until use.

**RCA-scWB.** Following electrophoretic separation and UV-mediated protein photocapture, PA-gel slides were washed to remove residual detergents and electrophoresis buffer. To reduce nonspecific adsorption of oligonucleotides and amplification reagents within the gel matrix, gels were first incubated in blocking buffer containing 3%BSA and 0.01mg/mL salmon sperm DNA (Thermo Scientific) for 1 h at room temperature. Photocaptured target proteins were recognized using the corresponding antibody–DNA conjugates. Gels were incubated with antibody–DNA conjugates at 5 μg $mL^{-1}$ overnight at 4 °C, followed by three washes with TBST to remove unbound antibodies.

**Padlock probe hybridization and ligation.** The padlock probe was allowed to hybridize to the antibody-linked DNA barcode for ligation. Ligation was performed at 37 °C for 60 min using T4 DNA ligase in the corresponding ligation buffer containing 1mM ATP (Thermo Scientific). Following ligation, the gel was washed to remove excess unbound oligonucleotides and ligation reagents. Exonuclease I (Thermo Scientific) were applied to digest unligated padlock probes and other accessible linear DNA while preserving successfully circularized DNA templates. Gels were incubated with the exonuclease mixture at 37 °C for 60 min

**Rolling circle amplification.** An RCA primer complementary to the circularized padlock probe was introduced into the gel and allowed to hybridize for 40 min. After primer annealing, rolling circle amplification was initiated using phi29 DNA polymerase (Sigma-Aldrich) in the presence of 1 mM each dNTP (Thermo Scientific) under the manufacturer-recommended reaction-buffer conditions. RCA was performed at 37 °C for 120 min. During amplification, phi29 DNA polymerase (Thermo Scientific) extended the annealed primer continuously around the circularized padlock template, generating a long single-stranded DNA concatemer containing tandem repeats complementary to the fluorescent detection probe. The amplification reaction was terminated by incubation at 65 °C for 10 min, after which gels were washed to remove polymerase, residual nucleotides, and soluble amplification components.

**Fluorescent probe hybridization.** RCA products were detected using fluorescently labeled DNA probes complementary to the repeated sequence generated during RCA. Fluorescent probes were diluted to a final concentration of 800 nM in hybridization buffer containing 2× saline-sodium citrate (SSC, Sigma-Aldrich) and Tween-20. The RCA-amplified gels were incubated with fluorescent detection probes at 37 °C for 30 min under light-protected conditions. Following hybridization, gels were washed three times to remove excess fluorescent probes using 2× saline-sodium citrate (SSC, Sigma-Aldrich) and reduce nonspecific fluorescence. The gels were prepared for fluorescence imaging.

**Small extracellular vesicle isolation and characterization.** Small extracellular vesicles were isolated from conditioned medium collected from HN6 and HN12 cell cultures by differential centrifugation. Conditioned medium was first centrifuged at 3,000 × g for 15 min to remove residual cells, cell debris, and larger particulate material. The resulting supernatant was transferred to ultracentrifuge tubes without disturbing the pellet and centrifuged at 100,000 × g for 1 h. The sEV-containing pellet was resuspended in PBS buffer and used for subsequent RCAmp-scWB measurements. Particle-size distributions of the isolated sEV preparations were characterized by nano-flow cytometry, confirming that the recovered particles were within the expected size range for sEVs.

**Statistics and reproducibility.** Data are presented as mean ± SD unless otherwise indicated. Independent devices were treated as the experimental replicates for statistical analyses, whereas individual cells, protein bands, or small extracellular vesicles measured within each device were treated as subsampled observations and were not considered independent replicates. For pooled single-cell or single-sEV distributions, individual measurements from independent devices were combined for visualization, while device-level values were used to summarize replicate variability and for statistical comparisons.

Unless otherwise indicated, experiments were performed using three independent devices. For paired comparisons in which the same experimental replicate was measured under both conditions, statistical significance was assessed using a two-sided paired Student's t-test. Exact P values are reported in the corresponding figure legends or main text. No statistical comparisons were performed for conditions represented by fewer than three independent devices. A P value $< 0.05$ was considered statistically significant. Statistical analyses were performed in MATLAB (version 2025b, MathWorks, Natick, MA, USA). No statistical method was used to predetermine sample size. No data were excluded from the analyses. Experiments were not randomized, and investigators were not blinded to experimental conditions unless otherwise stated.

**Acknowledgments:** This work was funded by the National Institutes of Health (NIH) 1R01GM16324. The study was also supported in part by Georgia CTSA/Regenerative Engineering and Medicine (REM) Pilot Grants Award.

**Competing interests:** The authors declare no competing interests.

**Data availability:** All the data are available from the corresponding authors upon reasonable request.

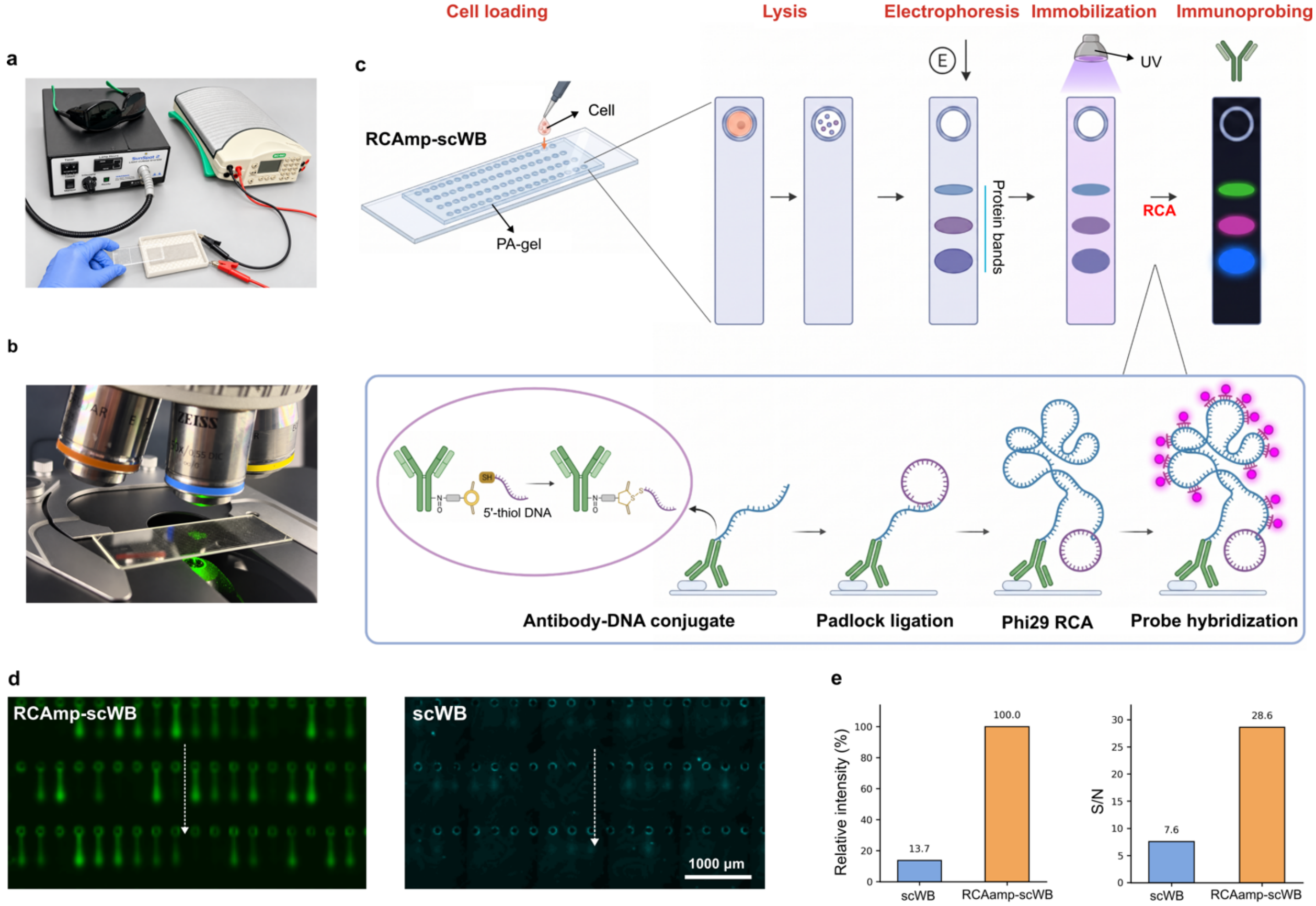

***Figure 1. Design and workflow of RCAmp-scWB.*** *(a) Photograph of the RCAmp-scWB electrophoresis setup. (b) Photograph of the optical measurement setup with the sample chip positioned on a motorized microscope stage. (c) Schematic of the RCAmp-scWB workflow. Single cells are loaded into microwells patterned in a polyacrylamide gel, followed by cell lysis, protein electrophoresis, photocapture, and immunoprobing with antibody–DNA conjugates. RCA is subsequently performed through padlock ligation, Phi29-mediated amplification, and fluorescent probe hybridization. The RCA detection chemistry uses antibody-DNA conjugates followed by padlock ligation, Phi29-mediated RCA, and fluorescent probe hybridization. (d) Representative fluorescence images of protein bands detected by RCAmp-scWB and conventional scWB. White arrows indicate the electrophoretic migration direction. Scale bar, 1000 µm. (e) Comparison of relative fluorescence intensity and signal-to-noise ratio (S/N) between conventional scWB and RCAmp-scWB.*

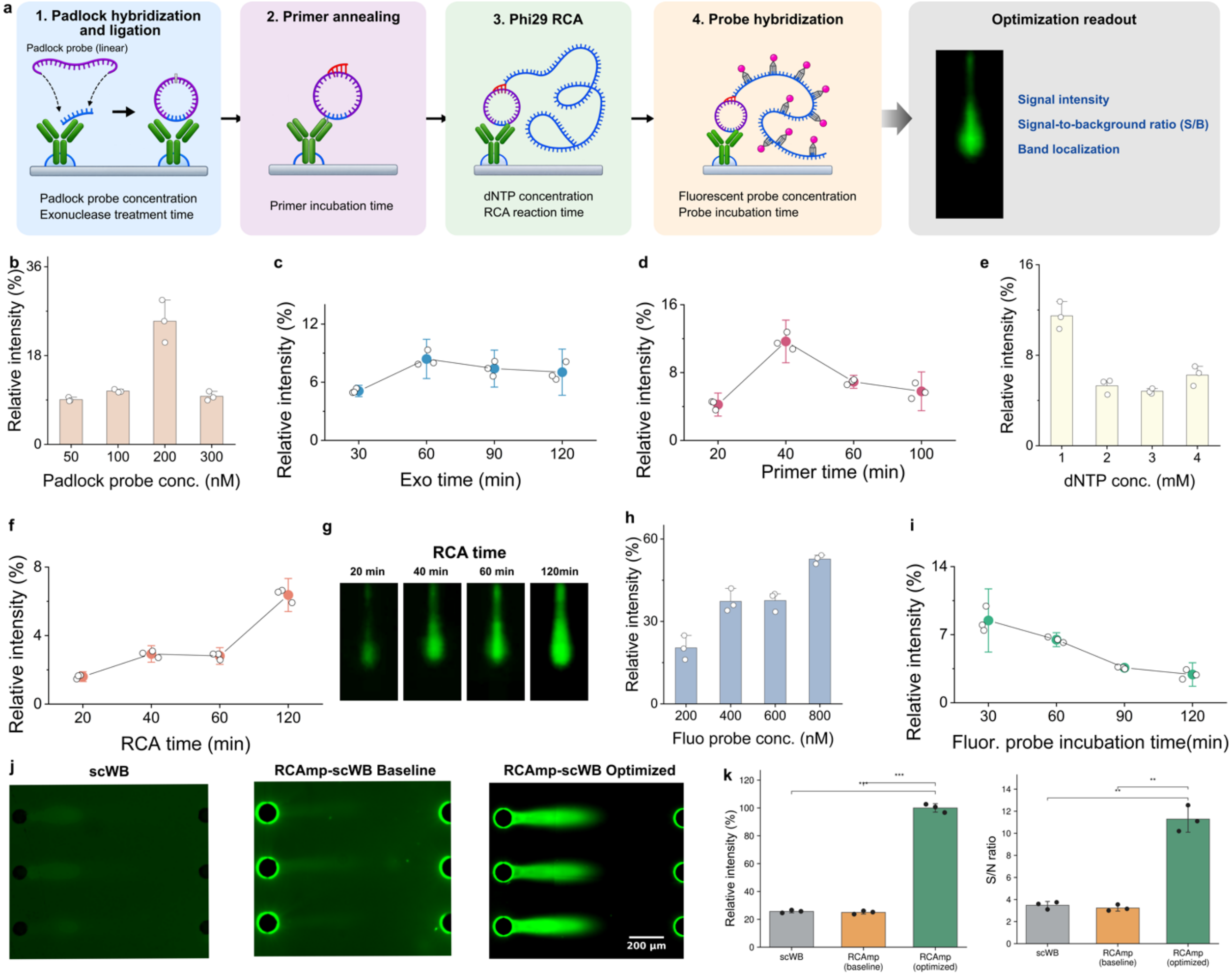

***Figure 2. Optimization of RCA conditions for RCAmp-scWB.*** *(a) Schematic of the RCA workflow and parameters evaluated for assay optimization. The workflow comprises padlock hybridization and ligation, primer annealing, Phi29-mediated RCA, and fluorescent probe hybridization. Signal intensity, signal-to-background ratio (S/B), and band localization were used as optimization readouts. (b) Relative fluorescence intensity at different padlock probe concentrations. (c) Relative fluorescence intensity at different exonuclease treatment times. (d) Relative fluorescence intensity at different primer incubation times. (e) Relative fluorescence intensity at different dNTP concentrations. (f) Relative fluorescence intensity at different RCA reaction times. (g) Representative fluorescence images of RCA-amplified protein bands at the indicated RCA reaction times. (h) Relative fluorescence intensity at different fluorescent probe concentrations. (i) Relative fluorescence intensity at different fluorescent probe incubation times. (j) Representative fluorescence images of protein bands detected by conventional scWB, baseline RCAmp-scWB, and optimized RCAmp-scWB. (k) Comparison of fluorescence intensity and signal-to-noise ratio (S/N) among conventional scWB, baseline RCAmp-scWB, and optimized RCAmp-scWB. Individual measurements are shown; data are mean ± SD (n = 3 independent experiments).*

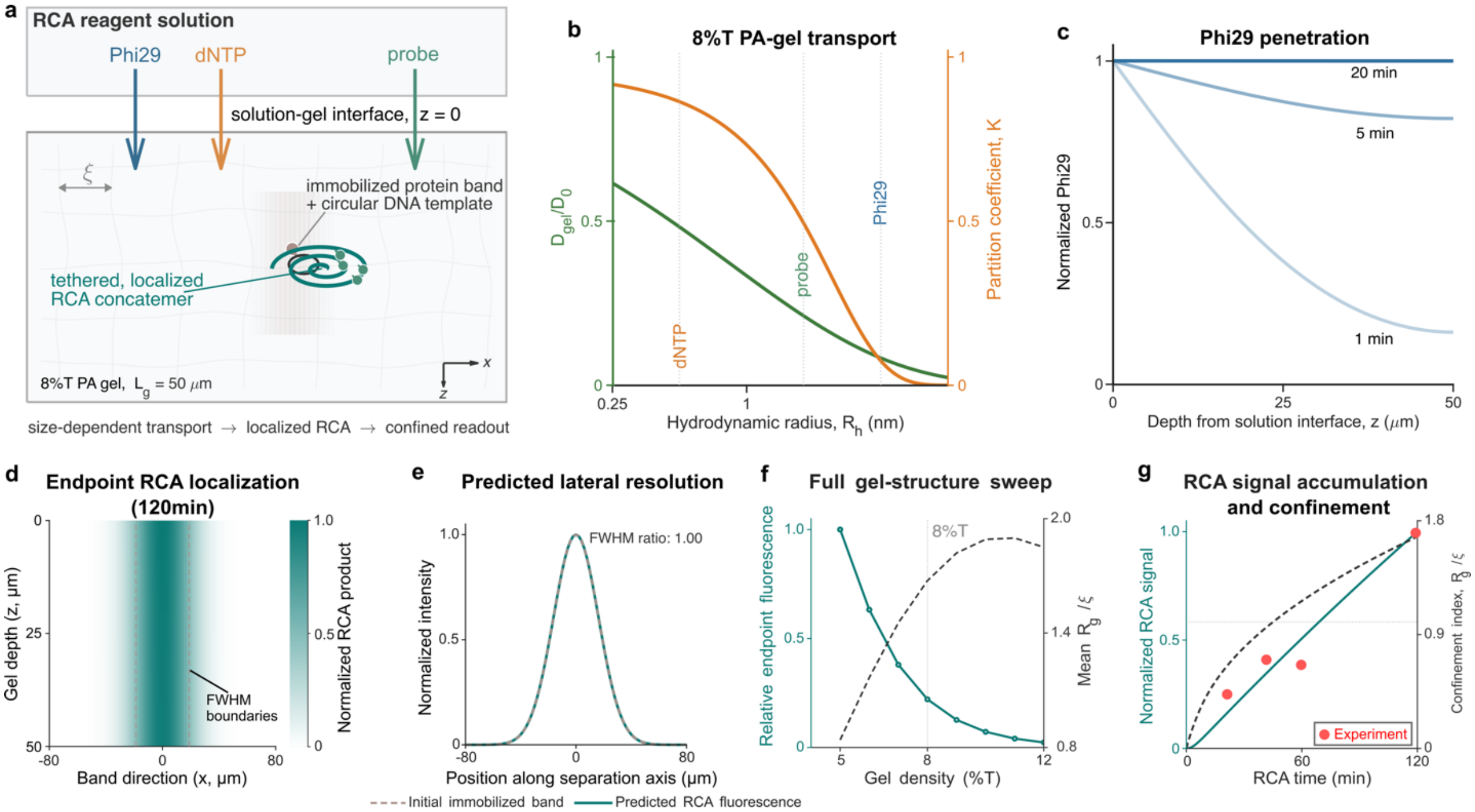

***Figure 3. Mechanistic modeling of gel-confined rolling-circle amplification.*** *(a) Two-dimensional model of molecular transport and localized RCA in an 8%T, 50-µm-thick polyacrylamide gel. Phi29 polymerase, dNTPs, and fluorescent probes enter the gel from the solution interface, whereas the RCA product remains tethered to the immobilized protein-associated circular DNA template. (b) Normalized in-gel diffusion coefficient, $D_{gel}/D_0$, and partition coefficient, K, as a function of molecular hydrodynamic radius. Vertical lines indicate the effective hydrodynamic radii of dNTPs, fluorescent probes, and Phi29 polymerase. (c) Simulated Phi29 concentration profiles across the gel depth after 1, 5, and 20 min. Concentrations are normalized to the gel-side equilibrium concentration. (d) Predicted spatial distribution of RCA product after 120 min. Dashed lines indicate the lateral full width at half maximum (FWHM) of the amplified band. (e) Predicted lateral fluorescence profiles of the initial immobilized band and RCA product under the tethered-product condition. The predicted FWHM ratio is 1.00. (f) Predicted effects of gel density on relative endpoint RCA fluorescence and the confinement index, $R_g/\zeta$, where $R_g$ is the effective radius of gyration of the RCA product and $\zeta$ is the gel mesh size. The experimental 8%T condition is indicated. (g) Predicted RCA signal accumulation and confinement during the 120-min reaction. Experimental mean fluorescence intensities from the RCA-time optimization experiment are shown for comparison. Simulated and experimental signals are independently normalized to their respective 120-min values.*

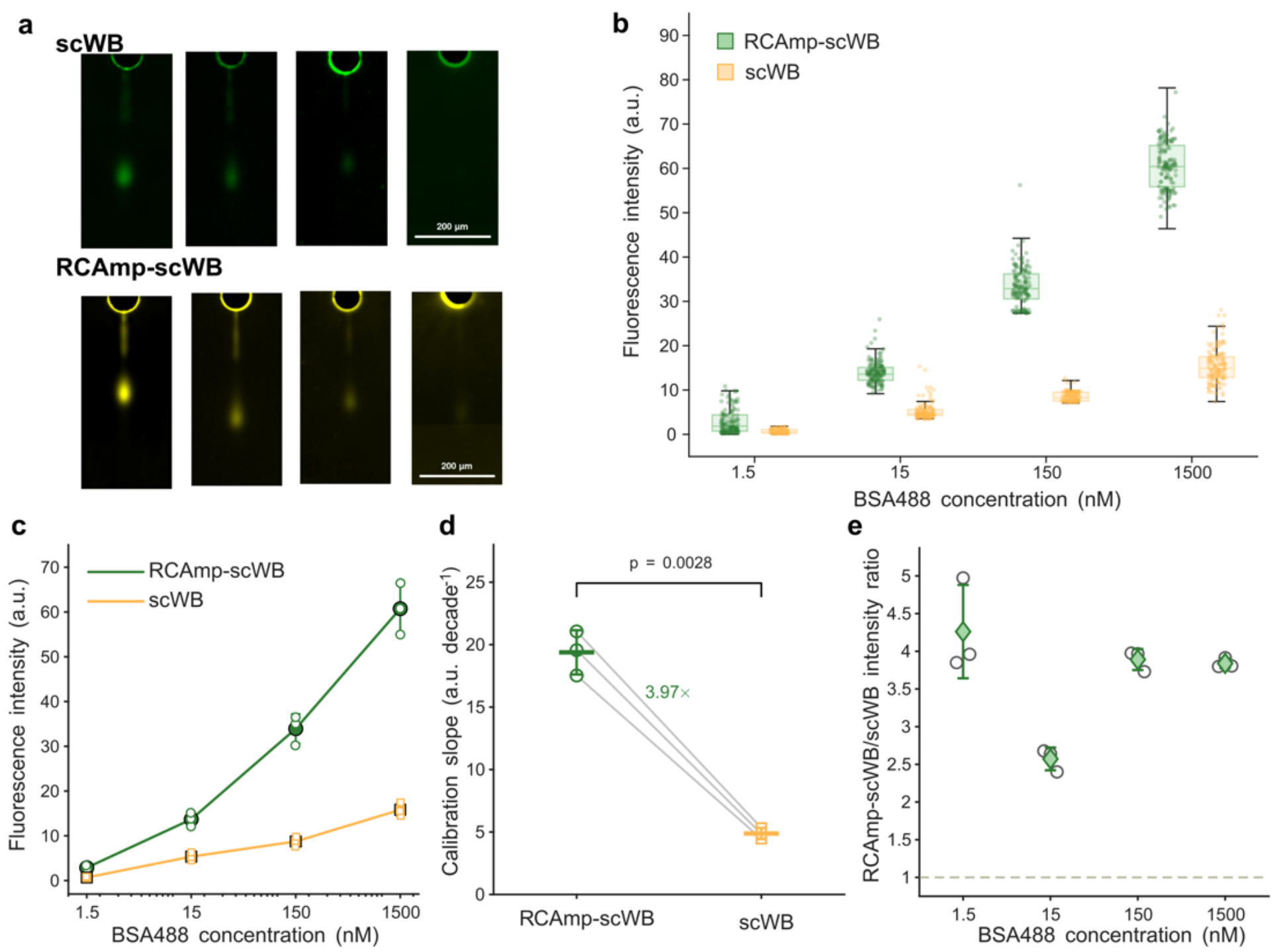

***Figure 4. Quantitative comparison of RCAmp-scWB and conventional scWB using purified BSA488.*** *(a) Representative fluorescence images of BSA488 detected by scWB and RCAmp-scWB at the indicated concentrations. (b) Distribution of fluorescence intensities measured by RCAmp-scWB and scWB across BSA488 concentrations from 1.5 to 1500 nM. Each point represents an individual protein band. (c) Mean fluorescence intensity as a function of BSA488 concentration for RCAmp-scWB and scWB. Individual experiment values and mean ± SD are shown (n = 3 independent experiments). (d) Calibration slopes obtained by linear regression of fluorescence intensity against log10 BSA488 concentration. Gray lines connect paired experiments. RCAmp-scWB yielded a 3.97-fold higher mean calibration slope than scWB. Statistical significance was determined using [statistical test], p = 0.0028. (e) RCAmp-scWB/scWB fluorescence intensity ratio across BSA488 concentrations. Individual experiment values and mean ± SD are shown (n = 3 independent experiments). The dashed line indicates an intensity ratio of 1.*

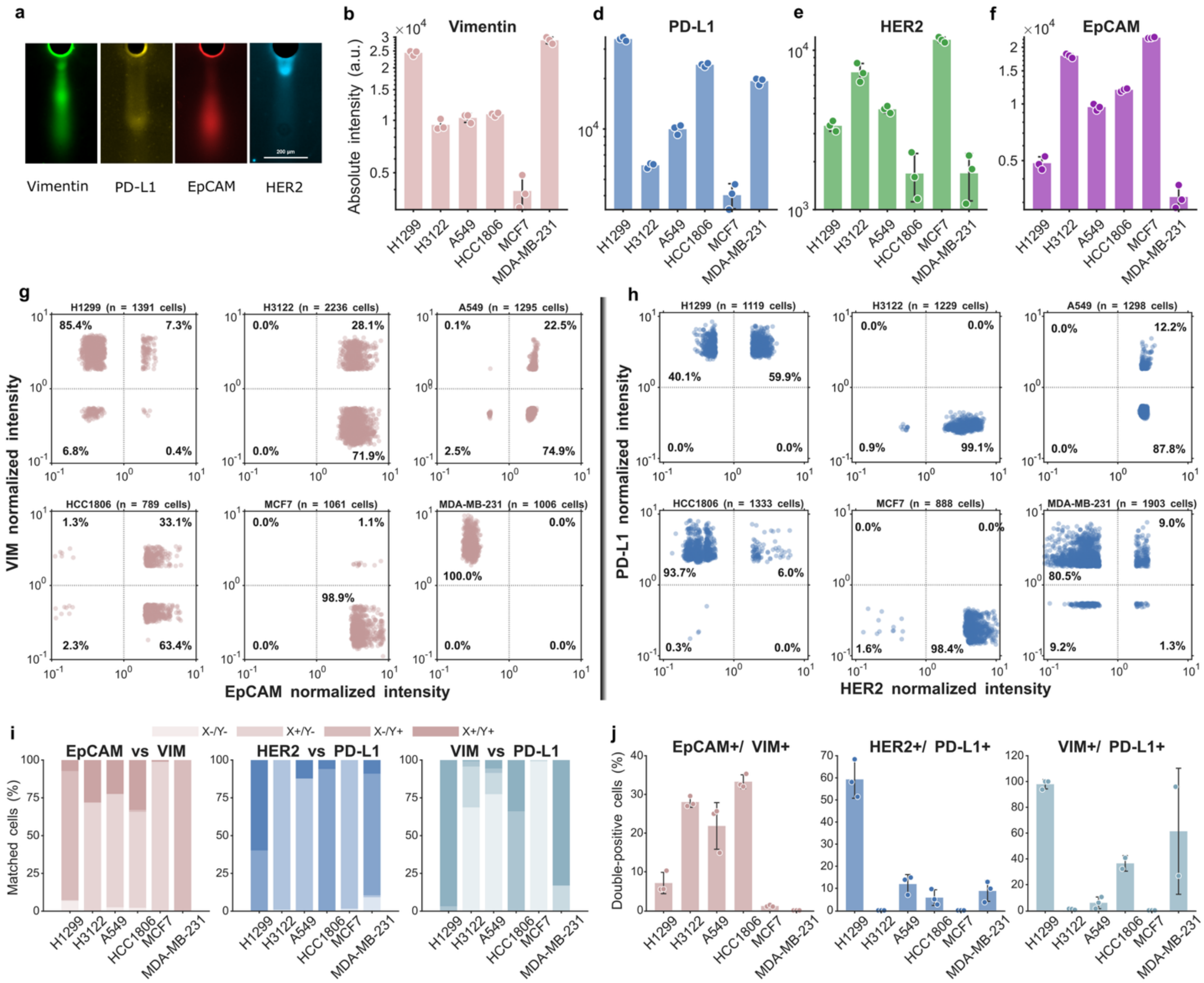

***Figure 5. Single-cell protein profiling across lung and breast cancer cell lines using RCAmp-scWB.*** *(a) Representative fluorescence images of Vimentin, PD-L1, HER2, and EpCAM detected by RCAmp-scWB. (b–e) Absolute fluorescence intensities of Vimentin, PD-L1, HER2, and EpCAM across H1299, H3122, A549, HCC1806, MCF7, and MDA-MB-231 cells. Individual device values and mean ± SD are shown (n = 3 independent devices). (f) Single-cell co-expression of Vimentin and EpCAM across the six cell lines. Dashed lines indicate quadrant thresholds, and percentages indicate the fraction of cells in each quadrant. (g) Single-cell co-expression of PD-L1 and HER2. Single-cell data in (f, g) were pooled from n = 3 independent devices. (h) Distribution of single-cell expression states for EpCAM/Vimentin, HER2/PD-L1, and Vimentin/PD-L1 across the six cell lines. For each protein pair, X and Y denote the proteins listed first and second, respectively. (i) Fraction of double-positive cells for EpCAM/Vimentin, HER2/PD-L1, and Vimentin/PD-L1. Individual device values and mean ± SD are shown (n = 3 independent devices).*

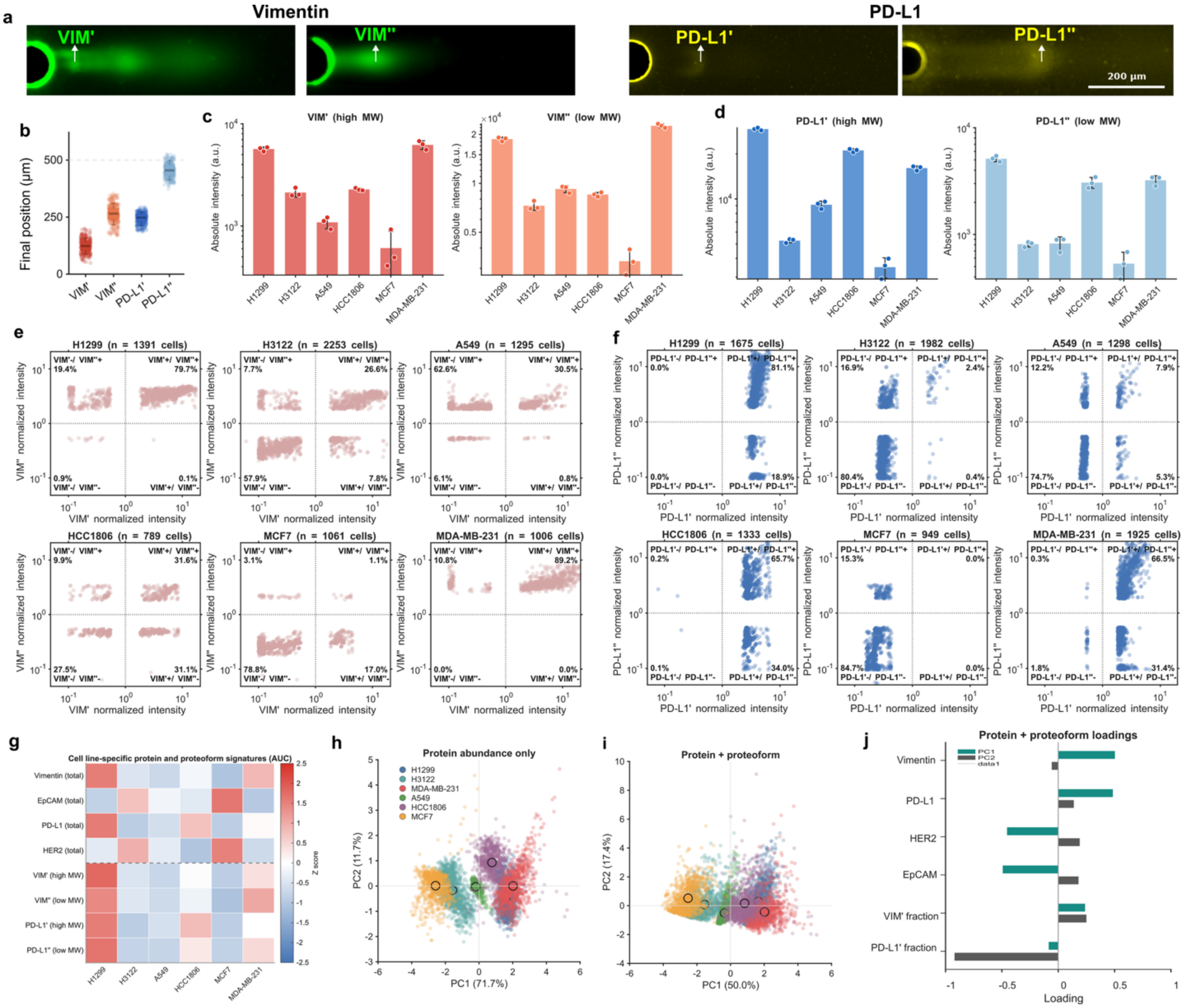

***Figure 6. Proteoform-resolved single-cell profiling of Vimentin and PD-L1 across cancer cell lines using RCAmp-scWB.*** *(a) Representative RCAmp-scWB fluorescence images showing resolved Vimentin and PD-L1 proteoforms. VIM' and PD-L1' denote the higher-molecular-mass species, whereas VIM'' and PD-L1'' denote the lower-molecular-mass species. (b) Distribution of electrophoretic migration positions of VIM', VIM'', PD-L1', and PD-L1''. (c, d) Absolute fluorescence intensities of the higher- and lower-molecular-mass Vimentin (c) and PD-L1 (d) proteoforms across H1299, H3122, A549, HCC1806, MCF7, and MDA-MB-231 cells. Individual device values and mean ± SD are shown (n = 3 independent devices). (e) Single-cell distributions of VIM' and VIM'' across the six cell lines. VIM' and VIM'' intensities are plotted on the x- and y-axes, respectively. (f) Single-cell distributions of PD-L1' and PD-L1'' across the six cell lines. Dashed lines in (e, f) indicate quadrant thresholds, and percentages indicate the fraction of cells in each quadrant. Single-cell data were pooled from n = 3 independent devices. (g) Heatmap of cell-line-specific protein and proteoform profiles across the six cell lines. Values were standardized as Z scores across cell lines. (h) Principal component analysis (PCA) based on total Vimentin, PD-L1, HER2, and EpCAM abundance. (i) PCA incorporating total protein abundance and the VIM' and PD-L1' proteoform fractions. (j) PC1 and PC2 loadings for the protein and proteoform features used in (i).*

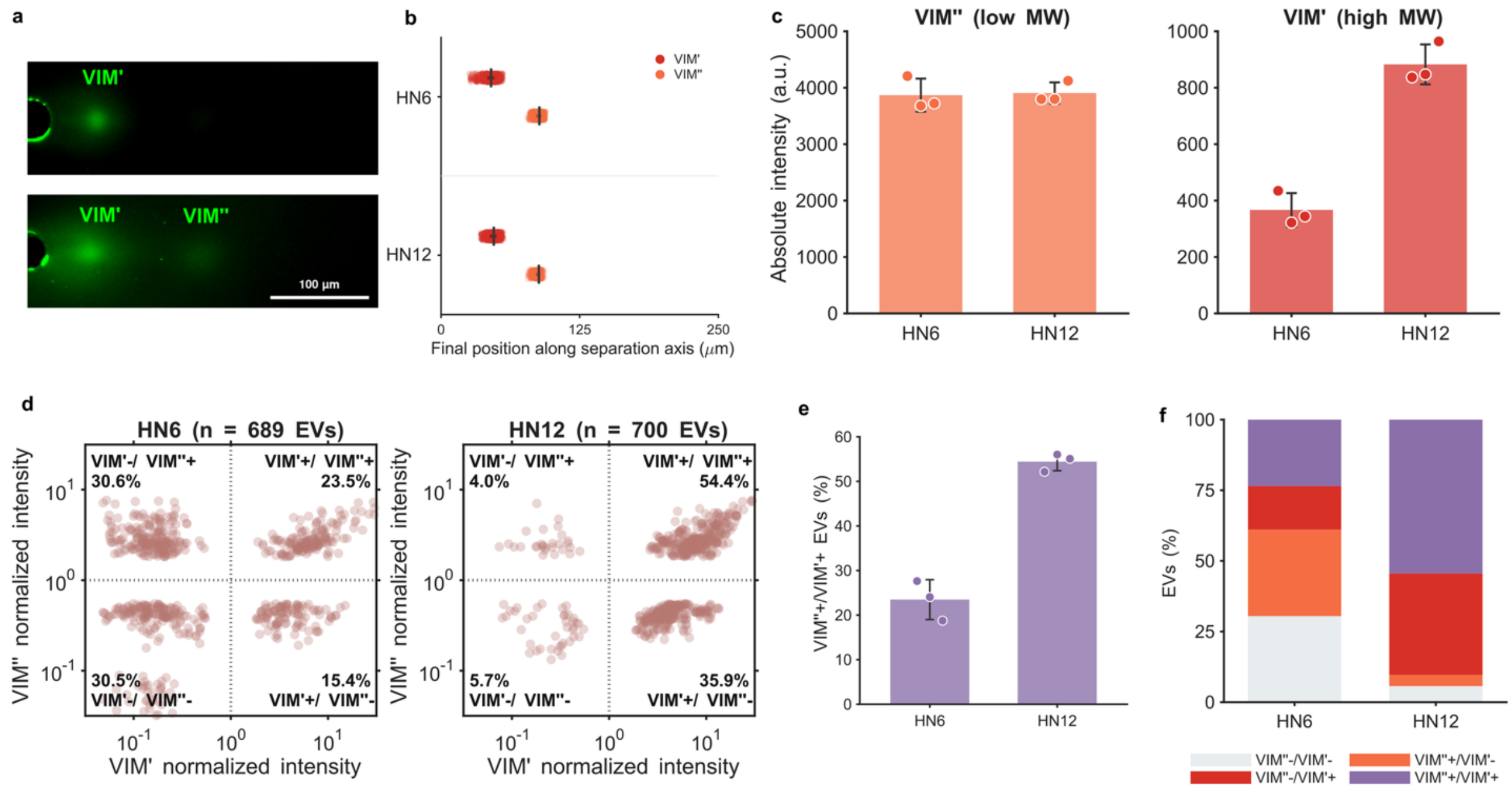

***Figure. 7 Single-sEV Vimentin proteoform profiling in head and neck cancer cell lines using RCAmp-scWB.*** *(a) Representative RCAmp-scWB fluorescence images of Vimentin proteoforms detected in small extracellular vesicle (sEV) samples. (b) Electrophoretic migration positions of VIM' and VIM'' in sEVs derived from HN6 and HN12 cells. VIM' and VIM'' denote the higher- and lower-molecular-mass species, respectively. (c) Absolute fluorescence intensities of VIM'' and VIM' in HN6- and HN12-derived sEVs. Individual device values and mean ± SD are shown (n = 3 independent devices). (d) Single-sEV distributions of VIM' and VIM'' in HN6- and HN12-derived sEVs. VIM' and VIM'' intensities are plotted on the x- and y-axes, respectively. Dashed lines indicate quadrant thresholds, and percentages indicate the fraction of sEVs in each quadrant. Single-sEV data were pooled from n = 3 independent devices (HN6, n = 689 sEVs; HN12, n = 700 sEVs). (e) Fraction of VIM'+/VIM''+ sEVs in HN6- and HN12-derived samples. Individual device values and mean ± SD are shown (n = 3 independent devices). (f) Distribution of Vimentin proteoform states in HN6- and HN12-derived sEVs.*